\documentclass[prl
,twocolumn%
,superscriptaddress%
,showpacs]{revtex4}
\usepackage{graphicx}%
\usepackage{dcolumn}
\usepackage{amsmath}
\makeatletter
\def\btt#1{\texttt{\@backslashchar#1}}%
\DeclareRobustCommand\bblash{\btt{\@backslashchar}}%
\makeatother

\begin{document}
%\setkeys{Gin}{draft=false}
%\setkeys{Gin}{draft=true}
\bibliographystyle{apsrev}
%%%%%%%%%%%%%%%%%%%%%%%%%%%%%%%%%%%%%%%%%%%%%%%%%%%%%%%%%%%%%%%%%%%%%%%%%
%\preprint{cond-mat/0000000}
\title{Spin caloritronics in magnetic tunnel junctions: \textit{Ab initio} studies} 
\author{Michael Czerner}%
\affiliation{%
I. Physikalisches Institut, Justus Liebig University, Giessen, Germany
}
\author{Michael Bachmann}%
\affiliation{%
I. Physikalisches Institut, Justus Liebig University, Giessen, Germany
}
\author{Christian Heiliger}%
 \email{christian.heiliger@physik.uni-giessen.de}
\affiliation{%
I. Physikalisches Institut, Justus Liebig University, Giessen, Germany
}
%
% It is always \today, today, but you may specify any date with \date.
\date{\today}
\begin{abstract}
This Letter presents \textit{ab initio} calculations of the magneto-thermoelectric power (MTEP) and of the spin-Seebeck coefficient in MgO based tunnel junctions with Fe and Co leads. In addition, the normal thermopower is calculated and gives for pure Fe and Co an quantitative agreement with experiments. Consequently, the calculated values in tunnel junctions are a good estimation of upper limits. In particular, spin-Seebeck coefficients of more than $100 \mu V/K$ are possible. The MTEP ratio exceed several 1000\% and depends strongly on temperature. In the case of Fe leads the MTEP ratio diverges even to infinity at certain temperatures. The spin-Seebeck coefficient as a function of temperature shows a non-trivial dependence. For Fe/MgO/Fe even the sign of the coefficient changes with temperature.
\end{abstract}
\maketitle
%
%%-----------------------------------------------------------------------
%
% Introduction
%
The emerging research field of spin caloritronics \cite{bauer10} combines the spin-dependent charge transport with energy or heat transport. In comparison to thermoelectrics the spin degree of freedom is considered as well. The influence of a temperature gradient on a spin-dependent current and vice versa was pointed out by Johnson and Silbsee \cite{silsbee87}. Since then a number of effects are discussed on the nanometer scale like thermal spin-transfer torque \cite{hatami07}, magneto-thermoelectric power (MTEP) in metallic multilayers \cite{gravier06}, thermally excited spin-currents \cite{tsyplyatyev06}, magneto-Peltier cooling \cite{hatami09}. 

Recently the spin-Seebeck effect was experimentally discovered by Uchida et al. \cite{uchida08} in a NiFe alloy. However, the interpretation of the measured effect is rather complicated. 
Thereby, the spins have different electrochemical potentials $\mu^\uparrow$ and $\mu^\downarrow$ due to a temperature gradient $\Delta T$ across  the sample. The spin-Seebeck coefficient is defined as
\begin{equation}
S_s=\frac{\mu^\uparrow-\mu^\downarrow}{\Delta T}
\label{eq:sS}
\end{equation}
There are in principle two effects that give rise to a spin voltage under an applied temperature gradient. The effect measured by Uchida et al. was recently explained by a spin pumping effect at the contact between the ferromagnet and the normal electrode \cite{xiao10}. 

The other effect is the analogue to the classical charge Seebeck effect. 
The origin of this effect is a different asymmetry of the density of states (DOS) around the Fermi energy in both spin channels. The asymmetry of the DOS is the main reason for a thermopower (or Seebeck voltage) in classical thermoelectrics. Introducing a magnetic material with different asymmetry in the DOS for both spins lead to different Seebeck coefficients for both spins $S^\uparrow$ and $S^\downarrow$. Both spin channels can be seen as a thermocouple leading to the spin-Seebeck coefficient 
\begin{equation}
S_s=S^\uparrow-S^\downarrow
\label{eq:Ss}
\end{equation}
For the classical thermopower a charge is spatially separated whereas as for the spin-Seebeck effect both spins are unequally occupied at the same position. Therefore, spin relaxation processes will destroy this effect if the sample size is larger then typical spin-diffusion lengths. Consequently, half metallic materials are promising. Nevertheless, the understanding of the spin caloritronic effects also for normal magnetic metals is of fundamental interest.
Surprisingly, Uchida et al. \cite{uchida08} measured a spin-Seebeck coefficient although the sample size is quite larger than the spin-diffusion length. Therefore, this measured effect has another origin as already pointed out above.

Due to these two different effects there is a confusion about the nomenclature. In particular, "`spin-Seebeck effect"' is used for both effects. The analogue of the charge Seebeck effect is given by the different Seebeck coefficients in both spin channels. Therefore, it is also possible to call this effect spin-dependent Seebeck effect. However, this is again confusing with respect to $S^\uparrow$ and $S^\downarrow$ which are the spin-dependent Seebeck coefficients. Therefore, throughout this letter we will use the nomenclature spin-Seebeck effect meaning the analogue to the charge Seebeck effect.

The effect of magneto-thermoelectric power (MTEP) is the dependence of the normal charge Seebeck coefficients on the relative magnetic orientation $\theta$ of both magnetic layers. The MTEP ratio is given by
\begin{equation}
\frac{S(0^\circ)-S(\theta)}{\min (|S(0^\circ)|,|S(\theta)|)}
\label{eq:MTEP}
\end{equation}
Gravier et al. \cite{gravier06} measured for $\theta=180^\circ$ a MTEP ratio of about 30\% in all-metallic junctions.

In this letter we investigate the spin-Seebeck effect (SSE) and the magneto-thermoelectric power (MTEP) in magnetic tunnel junctions. Thereby, we use ballistic transport that is in particular without spin-diffusion effects. For MTEP this is only a minor approximation because for the thermoelectric power (charge Seebeck coefficient) the electric charges are spatially separated which makes this effect robust. In the case of the SSE spin flip scattering destroys the effect leading to a vanishing spin-voltage if the sample size is larger than the spin diffusion length. Consequently, our investigations aim to give an upper limit of what is possible. This means that our calculated spin-Seebeck coefficients are basically only valid next to the barrier.

For a large Seebeck coefficient a strong asymmetry within the density of states is advantageous. In addition, for the spin-Seebeck a strong asymmetry within the spin channels is necessary. The latter is fulfilled for MgO based tunnel junctions with Fe or Co leads that show a very high tunnel magneto resistance (TMR) ratio \cite{gradhand08}. In such junctions MgO acts as an symmetry filter having a large transmission probability for $\Delta_1$ states only. With respect to these states Fe or Co is half-metallic having $\Delta_1$ states only in the majority spin channel. Therefore, one can expect also a high spin-Seebeck effect in MgO based tunnel junctions. A disadvantage of Fe or Co leads is that they are only half-metallic with respect to specific states. This means that the spin diffusion length is rather small in comparison to real half-metals.

Our method for the transport calculations is based on the Green's function formalism implemented in the Korringa-Kohn-Rostoker method \cite{heiliger08}. In this method non-collinear alignment of the magnetic layers can be considered to calculate the transport properties at an arbitrary relative angle between the magnetizations of the leads. The potentials are calculated self-consistently within a supercell approach for the parallel alignment of the magnetic moments of the magnetic layers. Due to the relatively thick MgO barrier of 6 monolayers both magnetic layers are decoupled. Therefore, the other magnetic orientations are obtained by rotating the potentials of the parallel alignment without an additional self-consistent cycle. For the calculation of the energy dependent transmission probability semi-infinite leads are considered by self energies. For both calculations the atomic sphere approximation is used and the cut-off for the angular momentum is 3. The energy dependent transmission probability $T(E)$ is used to calculate the moments
\begin{equation}
L_n=\frac{2}{h} \int T(E) (E-\mu)^n (-\partial_E f(E,\mu,T)) dE
\label{eq:L_n}
\end{equation}
where $f(E,\mu,T)$ is the Fermi occupation function at a given energy $E$, electrochemical potential $\mu$, and temperature $T$. The conductance $G$ and the Seebeck coefficient $S$ are given by \cite{ouyang09}
\begin{equation}
G=e^2 L_0 \ \ \ \ \ \ S=-\frac{1}{e T} \frac{L_1}{L_0}
\label{eq:G_S}
\end{equation}
For a better convergence with respect to the energy mesh we apply a very small bias voltage of $1meV$ to avoid sharp resonances in $T(E)$.
By using spin-dependent transmission probabilities $T^\uparrow(E)$ and $T^\downarrow(E)$ the spin-dependent Seebeck coefficients $S^\uparrow$ and $S^\downarrow$ are calculated. Eventually we use these spin-dependent Seebeck coefficients to obtain the spin-Seebeck coefficient using Eq. (\ref{eq:Ss}). By using $T(E)=T^\uparrow(E)+T^\downarrow(E)$ we calculate the charge Seebeck coefficient. The temperature dependence is included in the occupation function only.

\begin{figure}
\includegraphics[width=0.85 \linewidth]{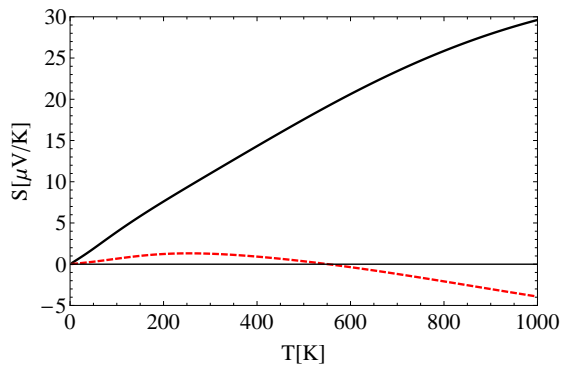}
\caption{
(Color online) Seebeck coefficient as a function of temperature for pure Fe (black, solid line) and Co (red, dashed line).
}
\label{S_pure}
\end{figure}

First we calculate the Seebeck coefficient for pure Fe and Co, where Co has the same bcc structure like Fe. 
Fig.~\ref{S_pure} shows the calculated results as a function of temperature. Experimental values are in the $\mu V /K$ range for pure Fe \cite{blatt67} and for a NiFe alloy \cite{uchida08} which means that our results have the correct order of magnitude. However, the details of the temperature dependence of Seebeck coefficient in pure Fe \cite{blatt67} is quite different to ours. Origins of these discrepancies are the not known quality of the samples and, in particular for high temperatures, missing inelastic contribution within the theory. To stress this point we are only investigating the temperature dependence due to changes in the occupation function. Nevertheless, our method is suitable to calculate the Seebeck and consequently the (ideal) spin-Seebeck coefficients in the right order of magnitude.

\begin{figure}
\includegraphics[width=0.85 \linewidth]{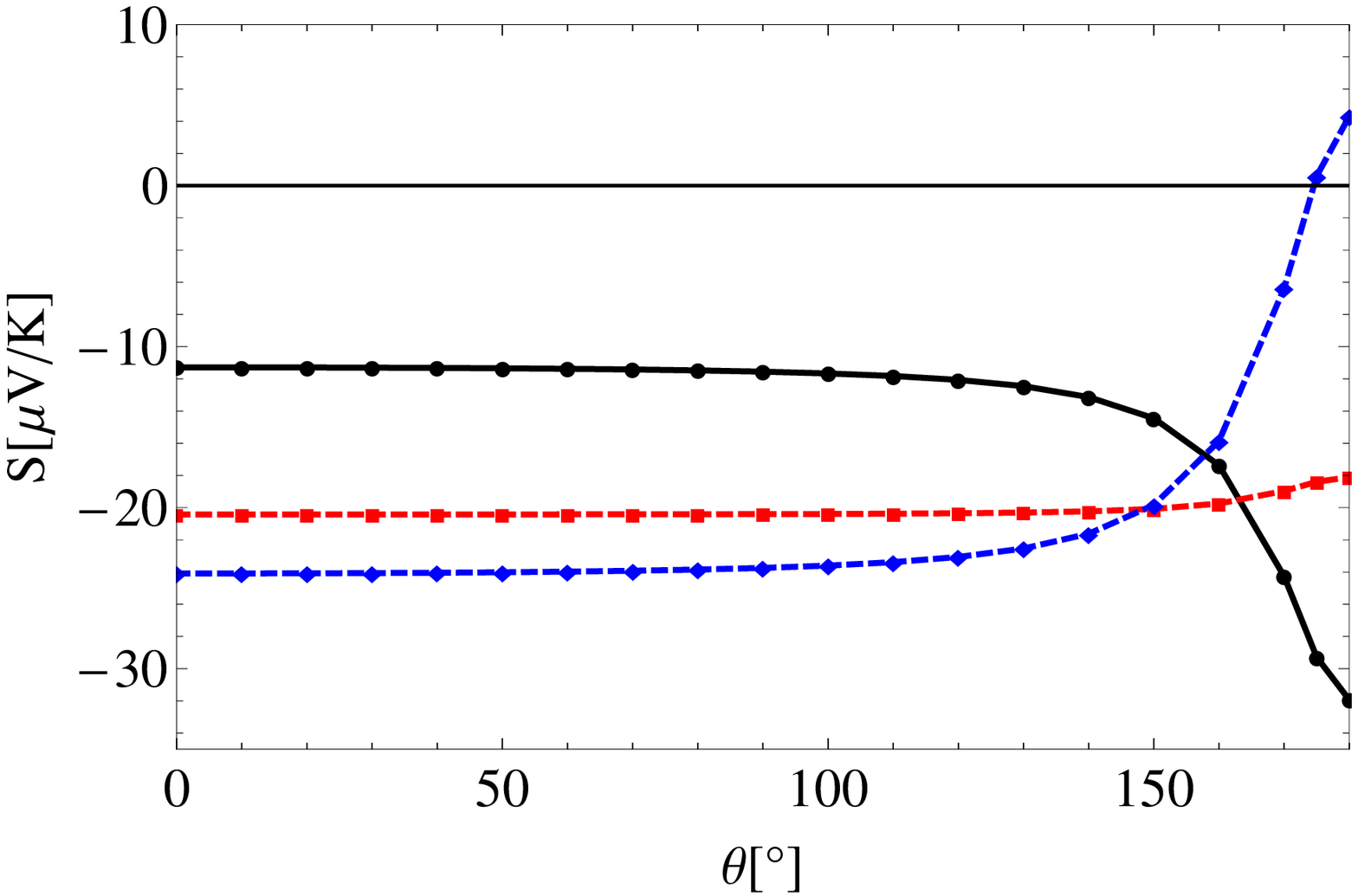}
\includegraphics[width=0.85 \linewidth]{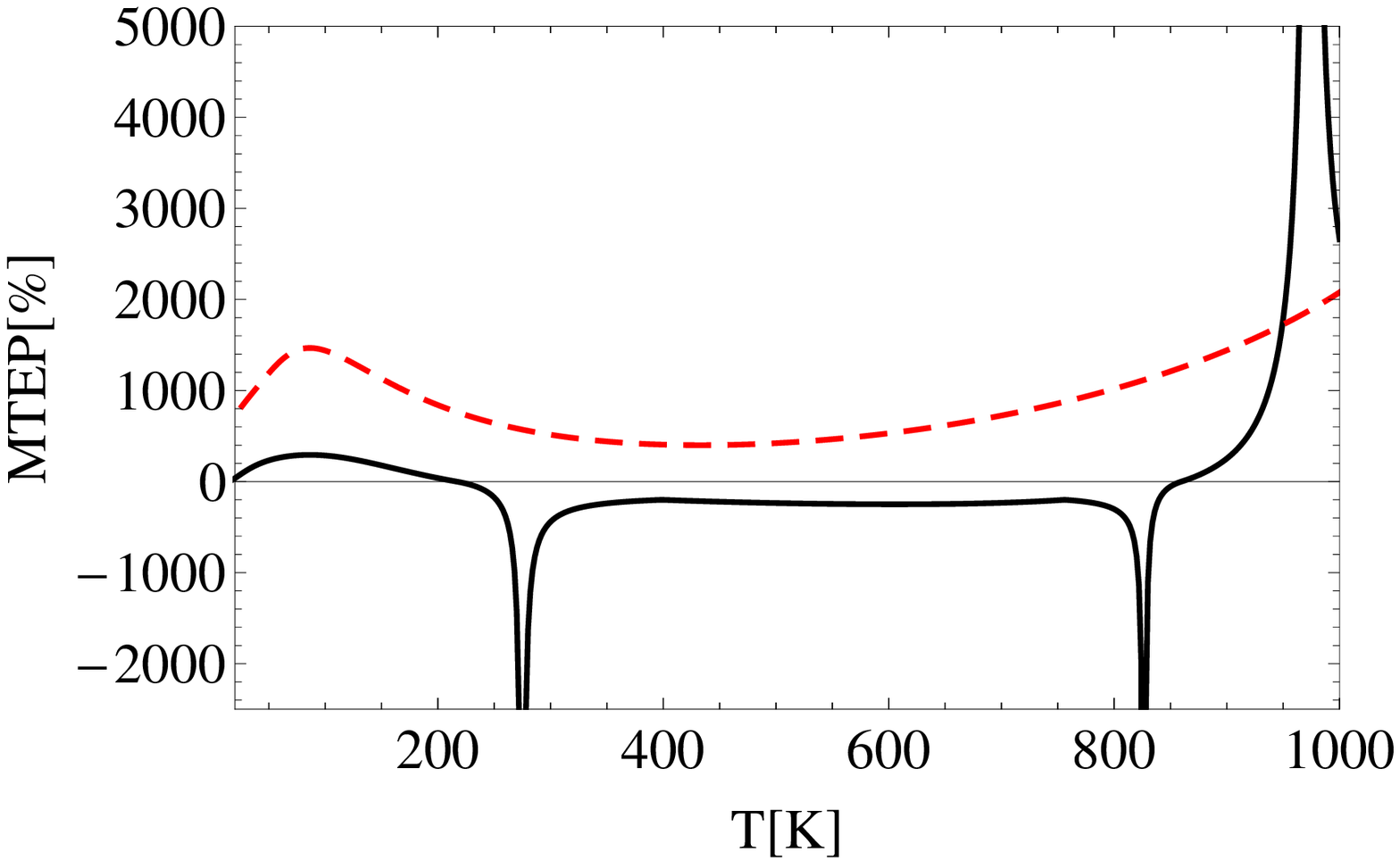}
\includegraphics[width=0.85 \linewidth]{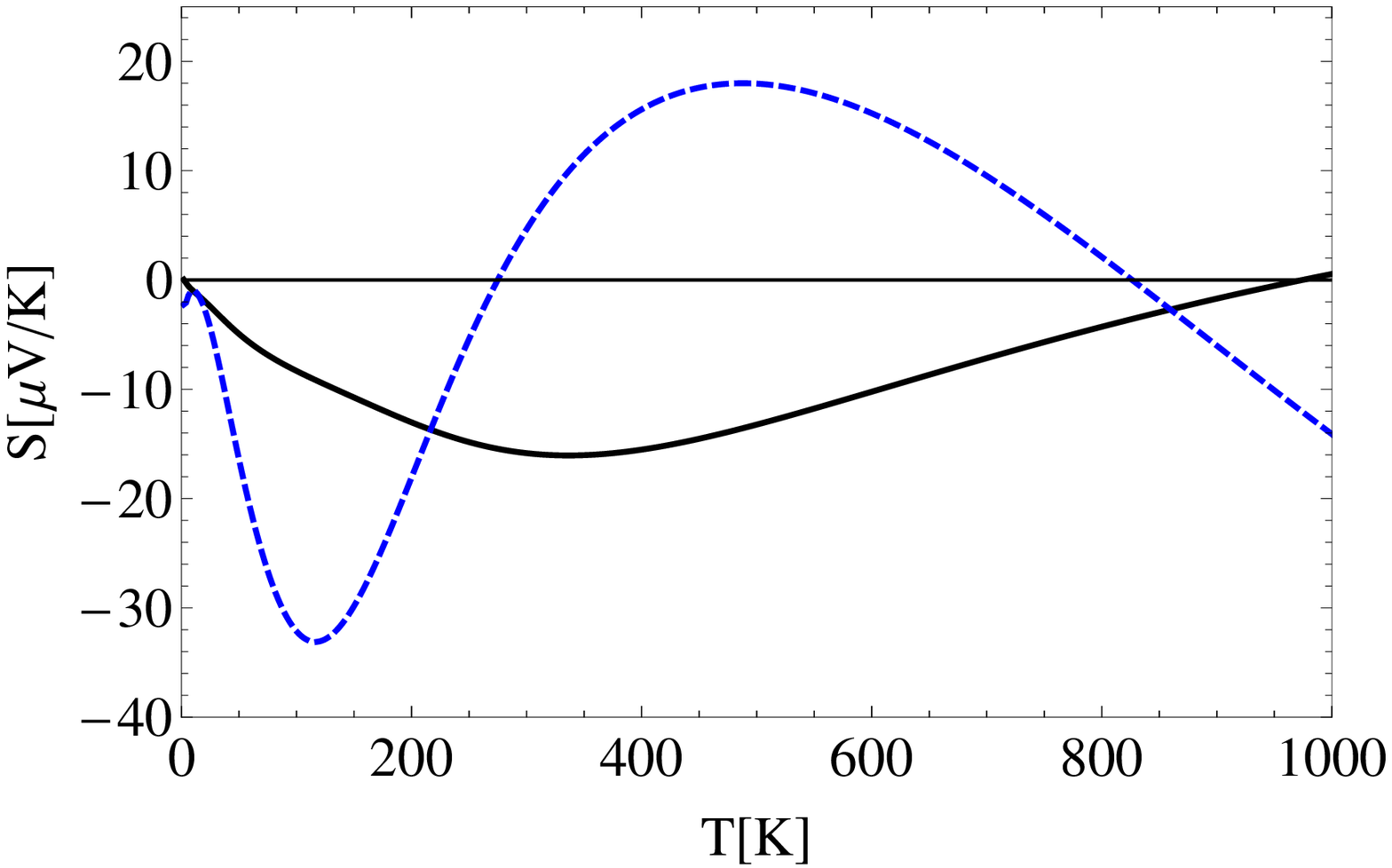}
\caption{ (Color online)
Upper panel: Seebeck coefficient of Fe/MgO/Fe as a function of the relative angle between the two magnetizations of the magnetic layers for different temperatures: 100K (black), 200K (red), and 300K (blue). 
Middle panel: MTEP ratio of Fe/MgO/Fe (black, solid line) and Co/MgO/Co (red, dashed line) as a function of temperature.
Lower panel: Seebeck coefficient of Fe/MgO/Fe as a function of temperature for parallel (black, solid line) and for anti-parallel (blue, dashed line) alignment of the magnetizations.
}
\label{MTEP}
\end{figure}

Next we investigate the MTEP in the tunnel junctions as a function of the relative magnetization of both magnetic layers to each other.
For this purpose, we look at symmetric tunnel junctions with Fe and Co leads, where Co has again the same structure like Fe. The magnetic layers are 20 monolayers, MgO has 6 monolayers, and the junction is connected to reservoirs represented by Cu in a bcc-Fe structure. The positions of the atoms are ideal to get only the influence of the magnetic material and not of different relaxations in addition. It is well known that the interface structure and therefore also relaxation at the interface can influence the transport characteristics. Therefore, we plan to do investigations of the influence of different interfaces on the spin-Seebeck coefficient and MTEP in the future. Fig.~\ref{MTEP} upper panel shows the Seebeck coefficient as a function of the relative angle of the magnetization for a temperature of 300 K. The angular dependence show an almost constant Seebeck coefficient up to about $120^\circ$. There is a drastic change at angles close to the anti-parallel alignment.
Fig.~\ref{MTEP} middle panel shows the temperature dependence of the MTEP ratio at anti-parallel alignment.
There is a huge MTEP effect that can be much larger than for all-metallic junctions which show an experimental value of about 30\% \cite{gravier06}. However, the temperature dependence is non-trivial including divergence at certain temperatures and change of sign. In addition, there is a large difference between the two magnetic materials.

Note that the Seebeck coefficient can be also negative. Therefore, it is not obvious which magnetic alignment causes the divergences of the MTEP ratio for the Fe/MgO/Fe junctions. Consequently, we show in Fig.~\ref{MTEP} lower panel the temperature dependence of the Seebeck coefficient for parallel and anti-parallel alignment. This viewgraph shows that the first two divergences for negative MTEP ratio are caused by a vanishing Seebeck coefficient for anti-parallel alignment. In contrast, the divergence at high temperature is due to a vanishing Seebeck coefficient in the parallel alignment. 

\begin{figure}
\includegraphics[width=0.9 \linewidth]{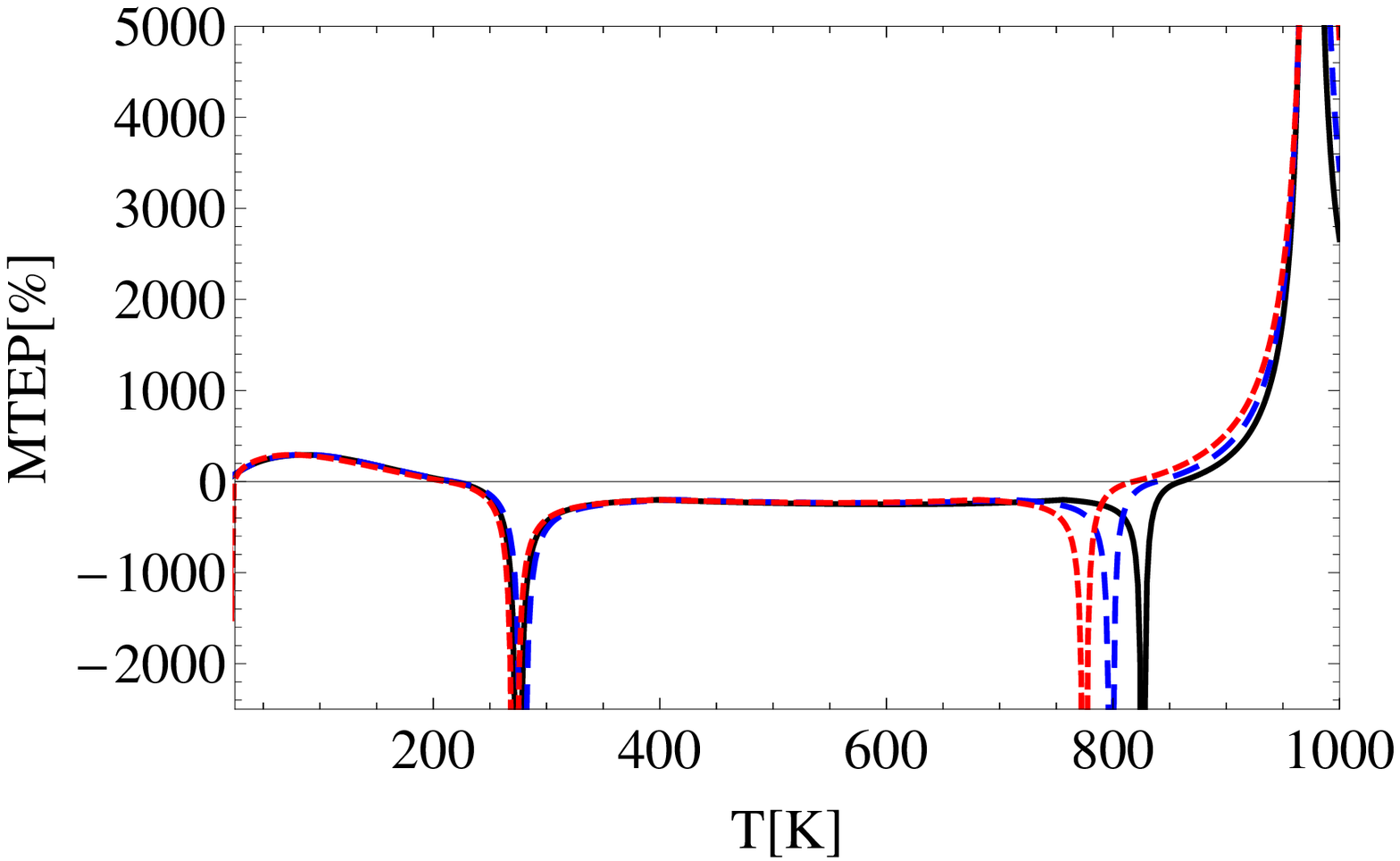}
\includegraphics[width=0.9 \linewidth]{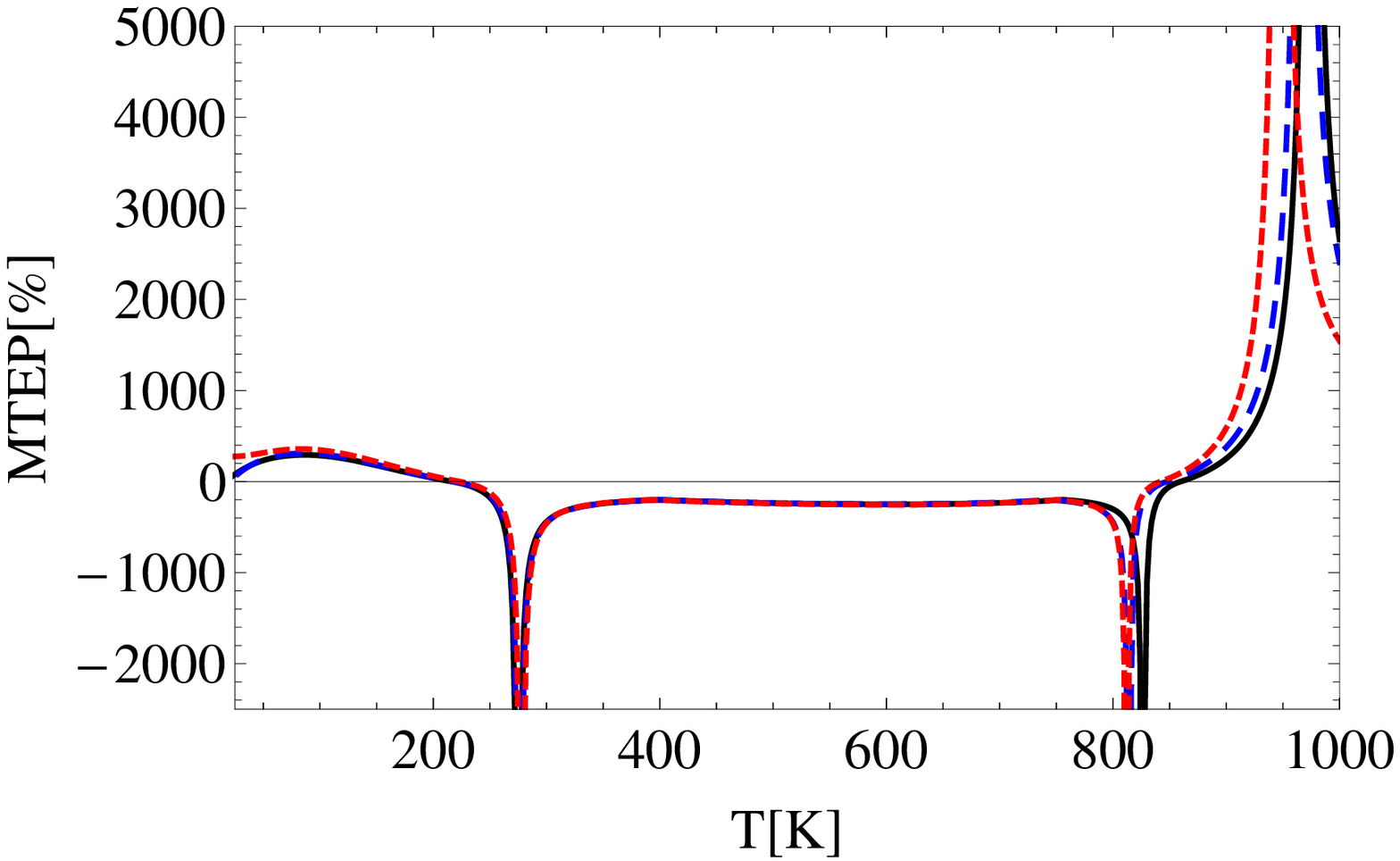}
\caption{ (Color online)
Upper panel: MTEP ratio of Fe/MgO/Fe for different k point meshes: 100x100 (10,000) k points (red, dashed line), 200x200 (40,000) k points (blue, dashed line), 400x400 (160,000) k points (black, solid line). The energy mesh for all cases has a distance of 0.68 meV between the energy points.
Lower panel: MTEP ratio of Fe/MgO/Fe for different energy meshes with a distance of 13.6 meV (red, dashed line), 3.4 meV (blue, dashed line), and 0.68 meV (black, solid line) between the energy points. In all cases we use a k point mesh of 400x400 (160,000) k points.
}
\label{conv}
\end{figure}

\begin{figure}
\includegraphics[width=0.9 \linewidth]{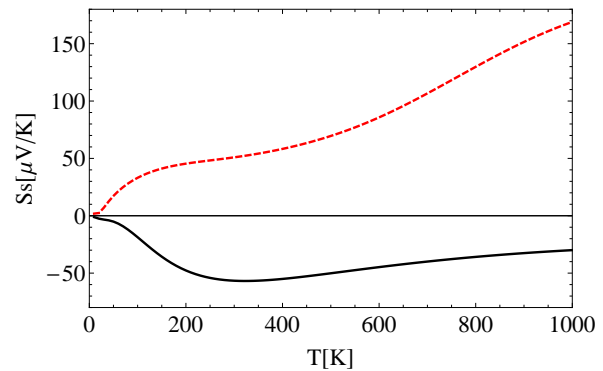}
\caption{ (Color online)
Spin dependent Seebeck coefficient of Fe/MgO/Fe (black, solid line) and Co/MgO/Co (red, dashed line) as a function of temperature.
}
\label{sS}
\end{figure}

\begin{figure}
\includegraphics[width=0.45 \linewidth]{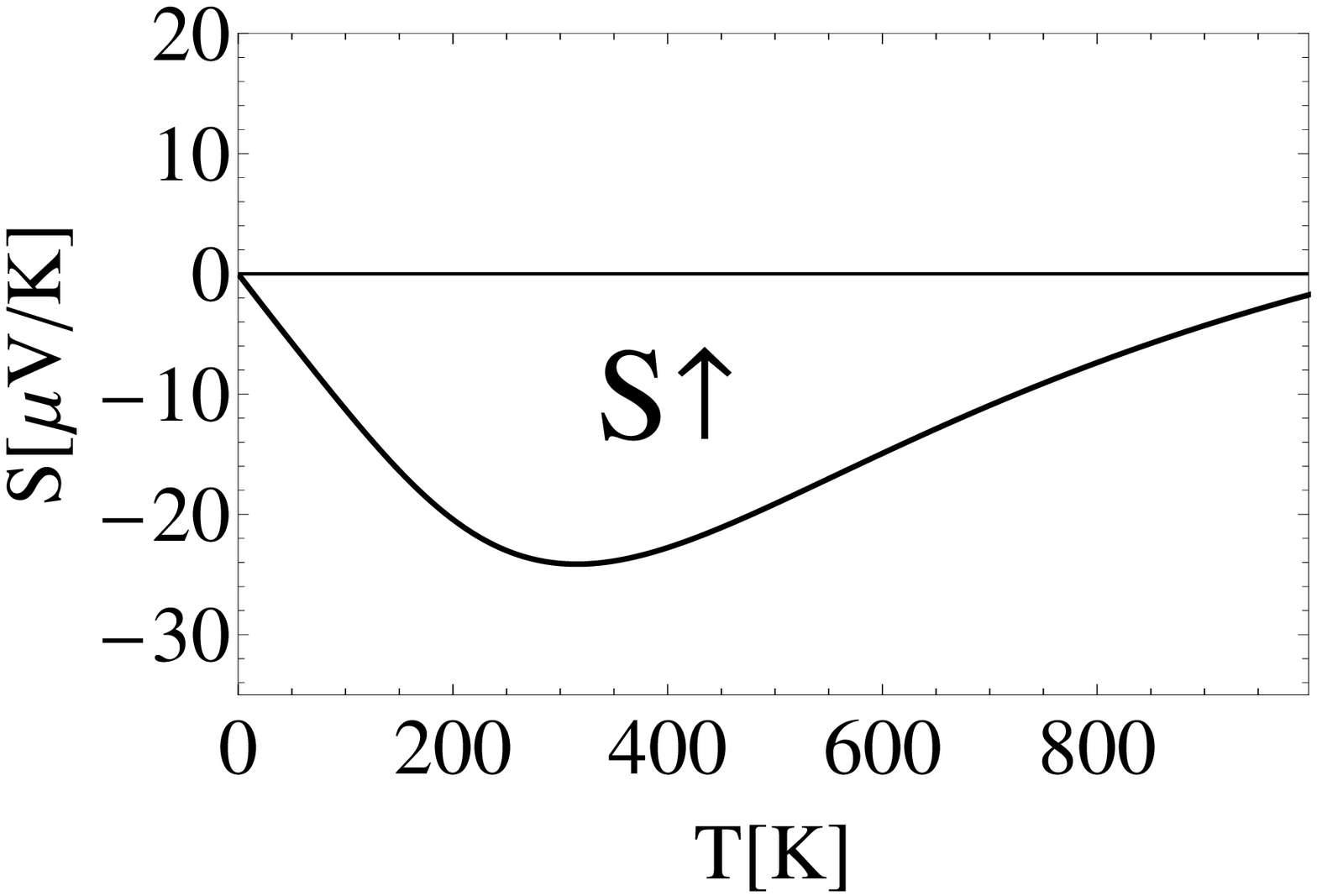}
\includegraphics[width=0.45 \linewidth]{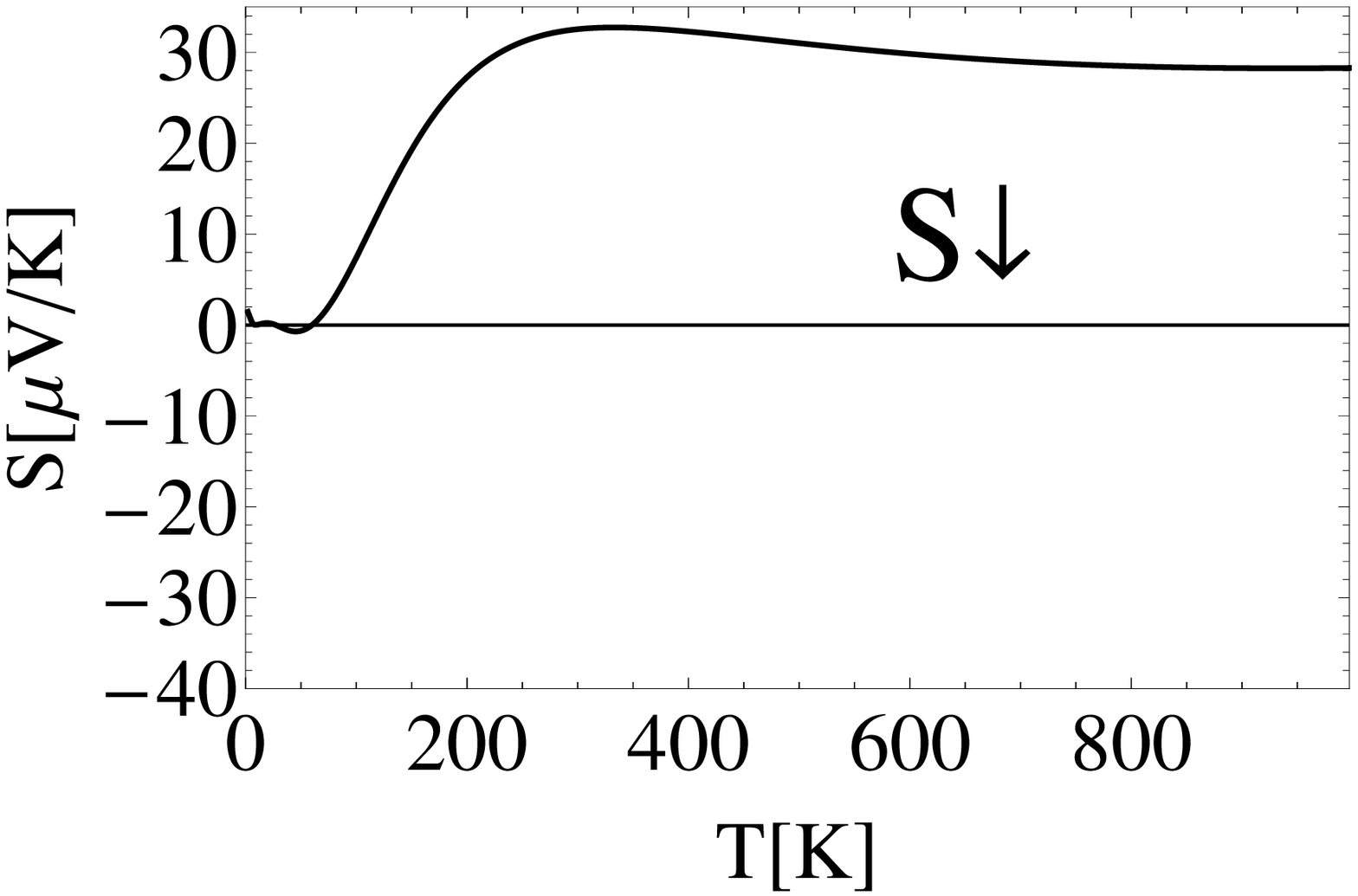}
\includegraphics[width=0.45 \linewidth]{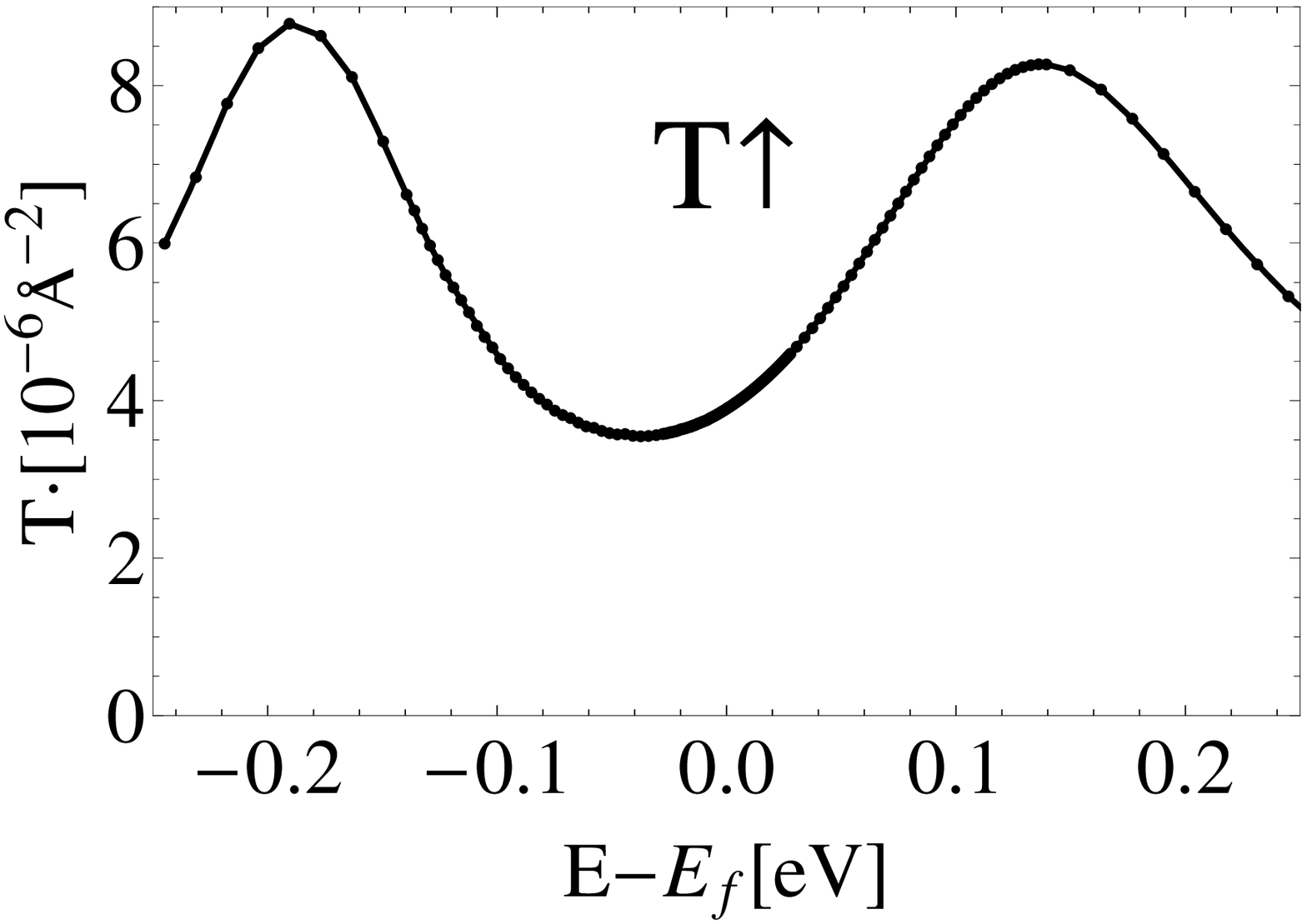}
\includegraphics[width=0.45 \linewidth]{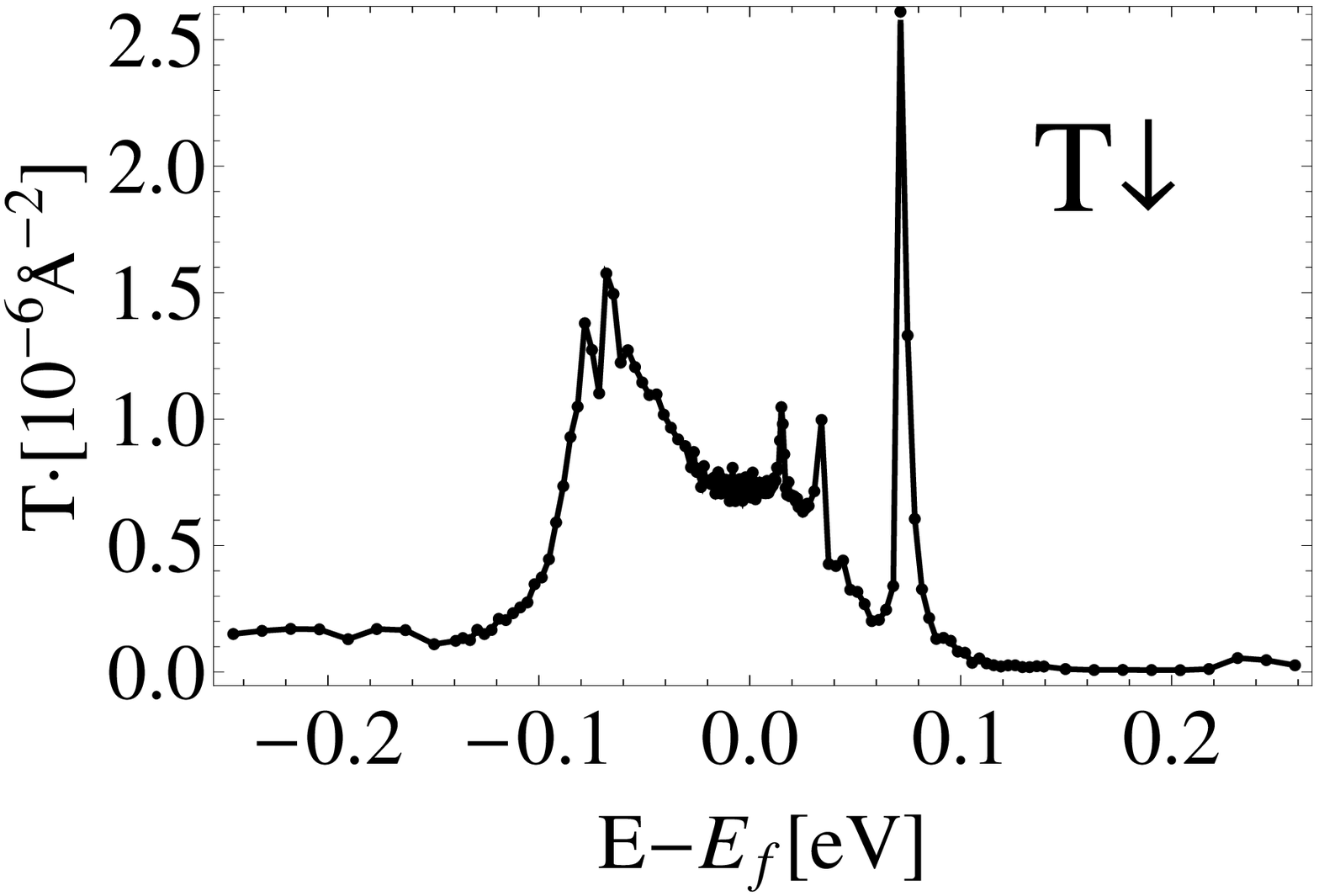}
\includegraphics[width=0.95 \linewidth]{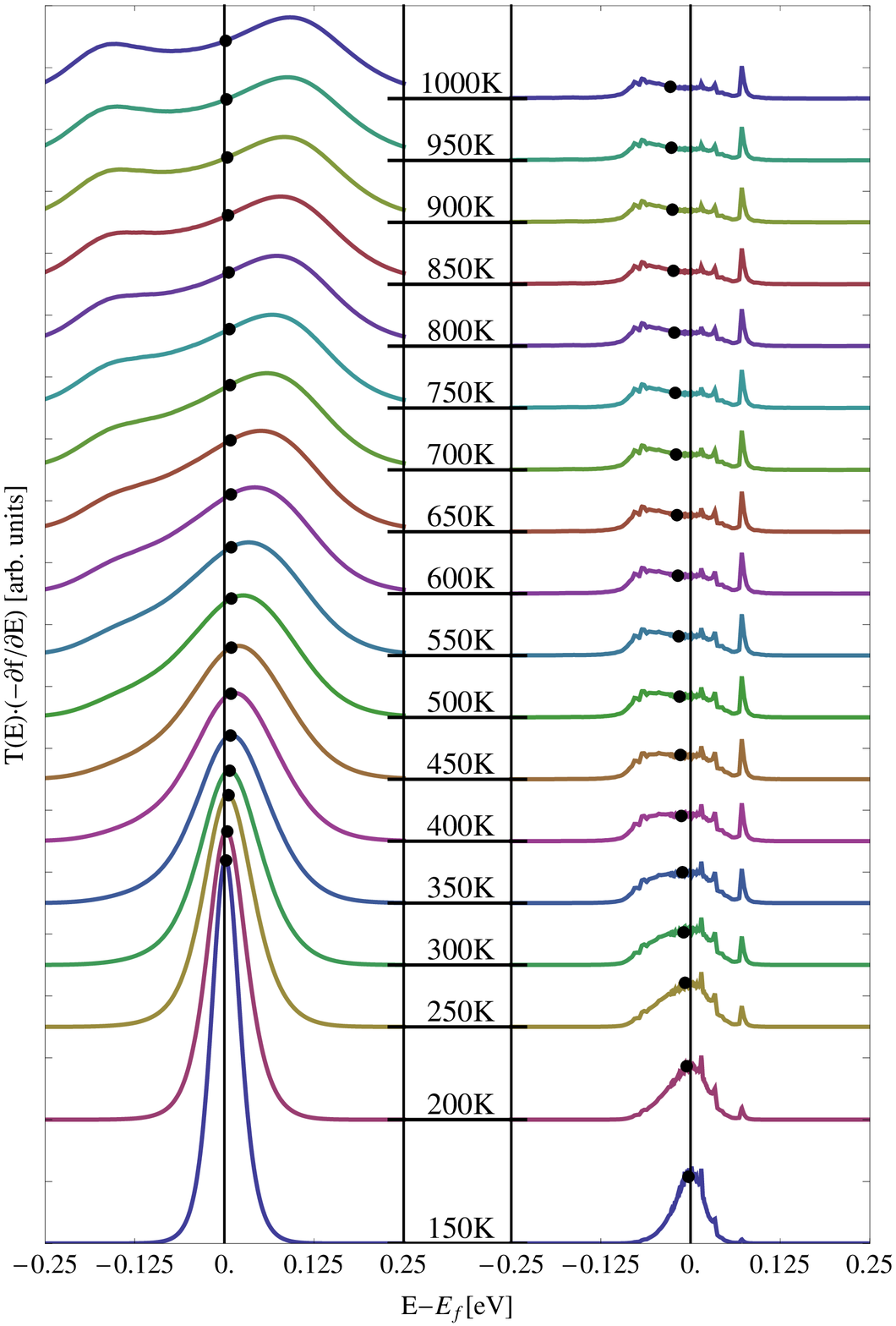}
\caption{ (Color online)
Top: Seebeck coefficient of Fe/MgO/Fe for the majority (left) and minority (right) spin.
Middle: Transmission probabilities as a function of energy for majority (left) and minority (right) spin.
Bottom: the integrand $T(E) \partial_E f(E,\mu,T)$ from Eq. (\ref{eq:L_n}) as a function of energy for majority (left) and minority (right) spin at different temperatures. The dot shows the center of mass.
}
\label{T_E}
\end{figure}

For magnetic tunnel junctions the calculation of transport parameters can be rather tedious due to the rich structure of $T(E)$ around the Fermi level. Even for one particular energy for $T(E)$ the k point mesh for the integration within the first Brillouin zone has to be very dense typically tens of thousands \cite{waldron07}. Consequently, convergence studies with respect to the number of k points and the number of energy points have to be carried out. The latter is in particular important for small temperatures. For this purpose Fig.~\ref{conv} shows the MTEP of Fe/MgO/Fe for different k point and energy meshes. The qualitative behavior of MTEP is basically independent of the number of k points. Only the position where the second divergence of the MTEP occurs changes slightly. The influence of the different energy meshes on the MTEP is similar. Main differences are at very small temperatures and the position of the third divergence of the MTEP. In both cases the position of the divergences changes their position due to the relative small slope of the Seebeck coefficients. A small change in the Seebeck coefficient shifts the point where the Seebeck coefficient vanishes and therefore the position of the divergence. Nevertheless the qualitative behavior is nearly unchanged. For Figs.~\ref{MTEP}, \ref{sS} and \ref{T_E} we actually use the larger k point mesh with 160,000 k points and the dense energy mesh with a distance between the energy points of 0.68 meV.

In Fig.~\ref{sS} we present the spin-Seebeck coefficients for Fe/MgO/Fe and Co/MgO/Co as a function of temperature. The absolute values are comparable to the classical Seebeck coefficients. However, note that these values are not robust and can be seen only as an upper limit. The temperature dependence for Fe/MgO/Fe is quite complicated with sign change of the slope with temperature.

The temperature dependencies of the MTEP and of the spin-Seebeck coefficient can be understood by looking at the energy dependent transmission probability. Features of these transmission probabilities on the other hand can be understood by looking at electronic structure at the interface between the magnetic material and the barrier \cite{heiliger06}. We will not further discuss the electronic states but we will discuss in Fig.~\ref{T_E} how $T(E)$ can explain the temperature dependence seen in Fig.~\ref{MTEP} and Fig.~\ref{sS}.

For this purpose we show in Fig.~\ref{T_E} the spin-dependent Seebeck coefficients and transmission probabilities for Fe/MgO/Fe. First, we start our discussion with the majority spin. In this case $T^\uparrow(E)$ is a smooth function showing two peaks one above and one below the Fermi level. The positions of the peaks are asymmetric to the Fermi level. Eq. (\ref{eq:L_n}) shows that the Seebeck coefficient is basically the center of the mass of $T(E) \partial_E f(E,\mu,T)$ divided by temperature. The contributing states within the integral are centered around the Fermi level and the width is increasing with increasing temperature. Consequently, starting from 0K the peak in the transmission above the Fermi level contribute to the Seebeck coefficient shifting the center of mass to higher energies which leads to the increase of the Seebeck coefficient. This is shown in the lower panel of Fig.~\ref{T_E}. When the temperature is large enough that the peak below the Fermi level contributes to the Seebeck coefficient the Seebeck coefficient starts to decrease. For very high temperatures both peaks contribute equally to the Seebeck coefficient leading to a center of mass closely to the Fermi level and a vanishing Seebeck coefficient. In a similar way the dependence of the Seebeck coefficient for the minority spin can be understood although $T^\downarrow(E)$ has a more complicated structure.

In summary, we calculated the spin-Seebeck coefficient and the magneto-thermoelectric power for MgO based tunnel junctions with Fe and Co leads. Spin-Seebeck values of up to $150 \mu V/K$ are possible which is similar to the value of the normal charge Seebeck coefficient. The calculated values can be seen as an upper limit of what is possible in experiments. Due to spin diffusion the spin-voltage will be strongly reduced with distance from the barrier. Nevertheless, our calculation shows what is the maximum possible difference in spin chemical potentials next to the barrier. Besides the absolute values we predict a non-trivial temperature dependence of the spin-Seebeck coefficient that changes drastically by going to other magnetic material. In particular, for Fe/MgO/Fe the sign of the slope of the Seebeck coefficient changes with temperature whereas for Co/MgO/Co the sign of the slope is the same for all temperatures. This means that different materials have different optimal working temperatures. The MTEP ratio can be several 1000\% in tunnel junctions. In particular, the non-trivial temperature dependence show even a divergence at certain temperatures. Consequently, in future work not only the material has to be analyzed in detail but also the temperature dependence.

We thank G.E.W. Bauer for useful discussion and acknowledge support from DFG SPP 1386 and DFG grant HE 5922/1-1.

%
%
%%---------------------------------------------------------------------------
%
% Produces the bibliography via BibTeX.
%\bibliography{lit}

%%%%%%%%%%%%%%%%%%%%%%%%%%%%%%%%%%%%%%%%%%%%%%%%%%%%%%%%%%%%%%%%%%%%%%%%%%%%%%%%

\end{document}